\documentclass{sig-alternate-05-2015}

\usepackage{newtxtext}
\usepackage{amssymb}
\usepackage{amsmath}
\usepackage{graphicx}
\usepackage[hide]{notes-alt}
\usepackage[ruled,vlined,linesnumberedhidden]{algorithm2e}
\usepackage{booktabs}
\usepackage{url}
\usepackage{listings}
\usepackage{color}
\usepackage[usenames,dvipsnames,svgnames,table]{xcolor}
\usepackage{dcolumn}
\usepackage{hyperref}
\usepackage{tabularx}

\newcolumntype{C}[1]{>{\centering}m{#1}}

\newdef{definition}{Definition}

\CopyrightYear{2017}
\setcopyright{acmcopyright}
\conferenceinfo{CHIIR '17,}{March 07-11, 2017, Oslo, Norway}
\isbn{978-1-4503-4677-1/17/03}\acmPrice{\$15.00}
\doi{http://dx.doi.org/10.1145/3020165.3022153}

\begin{document}
%
% --- Author Metadata here ---

% --- End of Author Metadata ---

% \title{History by Diversity: Helping Historians search News Archives

% \titlenote{(Produces the permission block, and
% copyright information). 
% \subtitle{[Working Title]
%\thanks{This work is partly funded by the European Research Council under ALEXANDRIA (ERC 339233)}

\title{Designing Search Tasks for Archive Search} 

 \numberofauthors{1}
 \author{
 \alignauthor
 Jaspreet Singh, Avishek Anand\\
        \affaddr{L3S Research Center, Leibniz Universit\"at Hannover.}\\
        \affaddr{Appelstr. 9a}\\
        \affaddr{30167 Hanover, Germany}\\
        \email{\{singh,anand\}@L3S.de}
 }

%\maketitle

\maketitle \begin{abstract} 

Longitudinal corpora like legal, corporate and newspaper archives are of immense value to
a variety of users, and time as an important factor 
strongly influences their search behavior in these archives. 
While many systems have been developed to support
users' temporal information needs, questions remain over how
users utilize these advances to satisfy their needs.
Analyzing their search behavior will provide us with novel insights
into search strategy, guide better interface and system design and highlight new problems for further research.
In this paper we propose a set of \emph{search tasks}, with varying complexity, that IIR researchers
can utilize to study user search behavior in archives. We discuss how we created and refined these tasks
as the result of a pilot study using a temporal search engine. 
We not only propose task descriptions but also pre and post-task evaluation mechanisms
that can be employed for a large-scale study (crowdsourcing).
Our initial findings show the viability of such tasks for investigating search behavior in archives.

%We also propose a new metric \textsc{Tia-Sbr} to evaluate the effectiveness of methods intending to solve it. 

\end{abstract}

% % A category with the (minimum) three required fields
% \category{H.4}{Information Systems Applications}{Miscellaneous}
% %A category including the fourth, optional field follows...
% \category{D.2.8}{Software Engineering}{Metrics}[complexity measures, performance measures]

% \terms{Theory}

% \keywords{ACM proceedings, \LaTeX, text tagging}

\vspace{-10pt}

\section{Introduction} % (fold)
\label{sec:introduction}

%Archives are growing increasingly important in a variety of fields. 
Text archives can be loosely defined as a collection of timestamped documents with the level of curation depending on the context.  Companies, governments, individuals, libraries and countless other organizations are continuously archiving documents for future reference and research. Such longitudinal corpora are not only important for historical research but also in retrospectively investigating scenarios in the legal and corporate world. While web search is geared towards popularity and recency, archive search is exclusively related to the past and time plays a vital role when searching.

%News articles and official reports, also often collected in archives, are used as sources for Wikipedia -- the largest collection of world knowledge on the web.

For instance, consider a historian who wants to know why chloroform was not used during childbirth in the Victorian age. 
Using the British Newspaper Archive's search engine~\cite{ukarchive}, she formulates her intent 
with the keywords \texttt{chloroform child birth 1800s}.

Ideally the system should help her uncover that
chloroform was first used in 1841 during child birth by Dr.James Chester who found that
chloroform did not pose any threat to mother and child. However
widespread usage in hospitals was delayed due to the opposition of
the church, till Queen Victoria used it in 1853
during the birth of Prince Leopold. The softening of the
church's stance then came in for much critique and debate in the late 1800s.

Researchers in the field of Temporal Information Retrieval (TIR) have developed novel retrieval models~\cite{croft,berberich2010language,histdiv}, query suggestion algorithms~\cite{Gupta2014}, indexing frameworks~\cite{Anand:2011} and exploratory search interfaces~\cite{expedition,neat,mishra2015expose} to directly address such scenarios and information needs. However, little work has been done to understand how users interact with specialized search engines and interfaces for specific search tasks. 
%Furthermore, only a fraction of this research has made its way to commercial systems used in libraries and large enterprises since little is known on how these methods improve search experience. 

We intend to address these concerns by devising a large-scale experiment that will allow us to investigate the following questions: (i) How do people interact with a temporal search system when searching for information in a text archive? (ii) Are the new retrieval models and interface elements useful for completing a search task? 
We hope that our work will shed light on user search strategy and help guide the design and development of such specialized search engines. In this short paper, we tackle the first hurdle of conducting such a study: devising  \emph{search tasks} for archive search. Research in Interactive IR has led to the creation of a plethora of standardized tasks but the information needs are generic and mostly intended for web search. On the other hand our tasks are tailored to a news archive -- The New York Times Newspaper Archive (1987 - 2007)~\cite{nyt}. We selected this particular archive over others for the following reasons: (i) easily available for a small fee (ii) clean XML storage format (iii) spans a large period of time (unlike web corpora). 

\begin{figure*}[t]  
%  \vspace{-12pt}
  \centering 
    \includegraphics[width=0.72\textwidth]{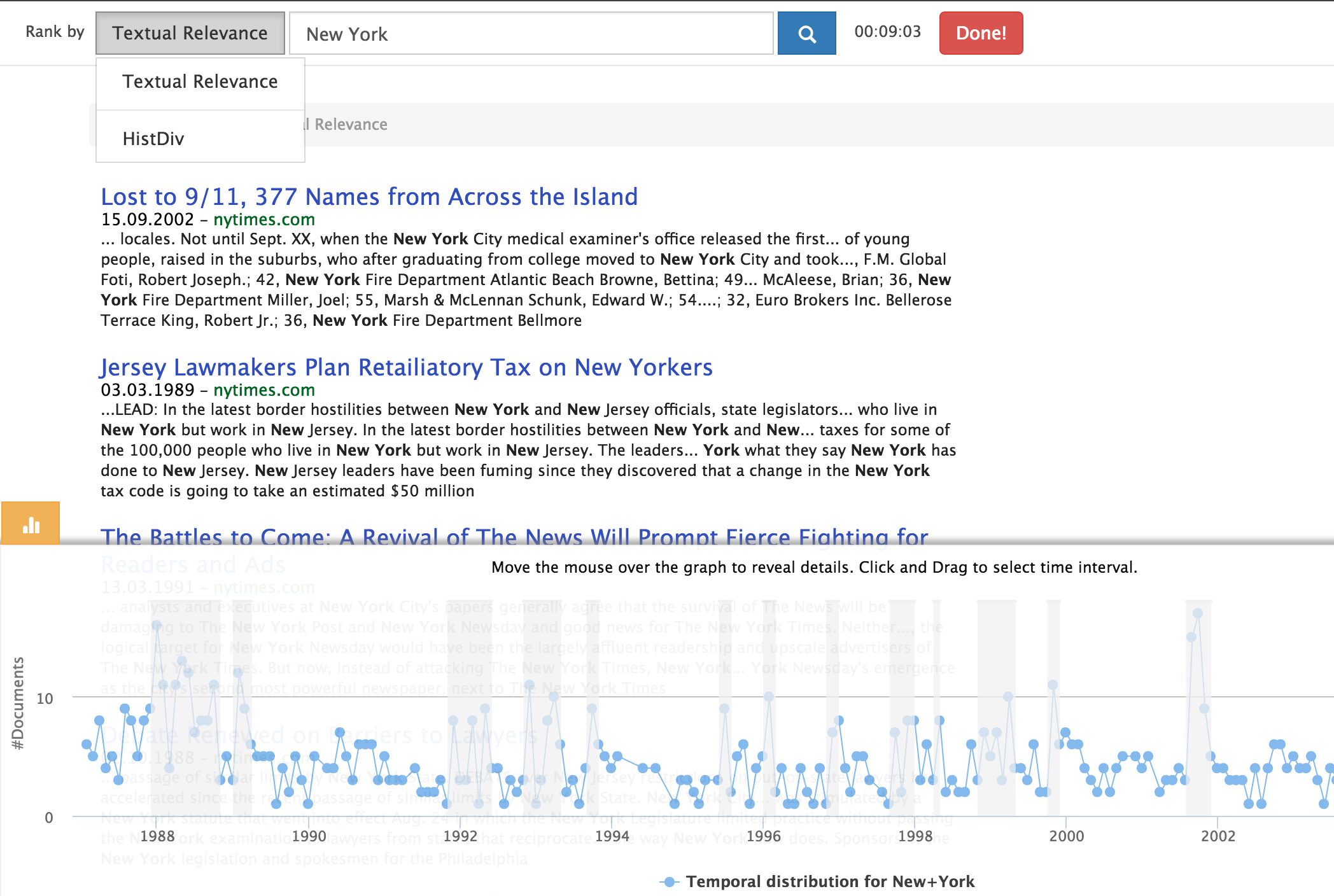} 
%\vspace{-15pt}
 \vspace{-6pt}
    \caption{Search User Interface of our Temporal Search Engine. The top of the user interface consists of a retrieval model selector with two choices: (i) Textual Relevance (default option, a language model with dirichlet smoothing) (ii) HistDiv (temporal diversification based ranking~\cite{histdiv}).  The bottom of the user interface is occupied by a timeline visualizing the publication history of a query. The timeline can be opened and closed by clicking on the yellow button. It is visible by default.}
    \label{fig:ui}
  \vspace{-9pt}
\end{figure*}

We created 27 search topics related to various domains using a cognitive complexity framework. We build directly on work by Kelly et. al.~\cite{kellyictir} but differ in the nature of the tasks and corpus. Additionally the tasks are designed with scale in mind, meaning the pre and post-tests suggested are simpler to evaluate.  We conducted a pilot study to confirm previous findings and refine our tasks before eventually using it for a future large-scale (crowd sourced) study. The participants were given access to a temporal search engine with two contrasting ranking methods: simple textual relevance (prevalent in commercial systems) and a diversification based approach designed specifically for archives~\cite{histdiv}. Additionally, they had the option of visualizing the publication history of the query on a timeline, so often used in state-of-the-art temporal exploratory search systems. Figure~\ref{fig:ui} shows the user interface of the search engine.

%Our results show that the tasks vary in complexity and the introduction of a timeline and alternate retrieval model are useful when working on these tasks.

% Chloroform was discovered in 1839 in England and first used
% in 1841 during child birth by \textsf{Dr.James Chester} who found that
% chloroform did not pose any threat to mother and child. However
% widespread usage in hospitals was delayed due to the opposition of
% the \textsf{church}, till \textsf{Queen Victoria} used it in 1853
% during the birth of \textsf{Prince Leopold}. The softening of the
% church's stance came in for much critique and debate subsequently.

% Temporal information needs have been characterized in two ways.  In \cite{}, the user's information need is linked to time either explicitly (e.g. olympics 2016) or implicitly (e.g. wtc bombing). Implicit temporal queries in addition can be ambiguous -- wtc bombing can refer to 9/11 or the '93 bombings. Alternatively, \cite{} consider the temporal ambiguity of a query to be distributed across 4 fixed classes -- past, present, future and atemporal. However in both cases, the information needs are defined for users of general web search engines. An archive is different from the web and people looking for information 

% section introduction (end)

\section{Background} % (fold)
\label{sec:background}

One of the major challenges for TIR, as outlined in~\cite{tir}, is to understand how people search when their information intent is inherently temporal. Temporal information needs are either explicit (e.g. olympics 2016) or implicit (e.g. wtc bombing)~\cite{campos_survey_2014}. Implicit temporal queries can be ambiguous -- wtc bombing can refer to 9/11 or the 1993 bombings. Alternatively, \cite{histdiv} introduces the concept of a historical query intent where the user wants an overview of a broad topic across time, implying that temporal diversity is inherent to the intent. Several retrieval models have been suggested for these special intents~\cite{berberich2010language,croft,histdiv} along with prototype systems that integrate them with timeline based visualizations in exploratory search interfaces~\cite{expedition,neat,mishra2015expose}. However we do not know how user search experience and behavior is affected when using these systems.

To study this, a set of simulated search tasks is necessary. \cite{wildemuth} describes a search task as a goal oriented activity carried out using search systems. Kelly et. al.~\cite{kellyictir} recently laid the groundwork for researchers to design their own search tasks based on Anderson and Krathwohl's taxonomy of educational objectives~\cite{krathwohl}. It consists of 6 types of cognitive processes of increasing complexity: remember, understand, apply, analyze, evaluate and create; as well as an associated target outcome.  For instance, the outcome of a remember task is a fact whereas the outcome of an analyze task is a prioritized list of facts. Previous attempts to create search tasks and queries for newspaper archives~\cite{braun2002information,Maslov:2006:ENQ:1135777.1135950} have either relied on experts and/or logs but fail to consider varying levels of complexity.
%Kelly et. al. also provide pre and post-search questionnaires that measure perceived task difficulty. 

%Our overall objective is to study search behavior by varying both task complexity and user interface features. 
To the best of our knowledge, no search tasks have been designed to specifically study user search behavior in archives. There exists work like~\cite{jarvelin2015information} dedicated to applying conventional IR techniques to historical archives, however, interactive IR remains relatively unexplored. While previous work has shown that more complex search tasks lead to increased user interaction with the search engine~\cite{liu2010search}, we wish to more specifically investigate how users interact with timeline visualizations and the ability to switch between retrieval models when completing these search tasks. 

%In the remainder of this paper we discuss the procedure used to create the search tasks, the interface of the temporal search engine and then the results of a pilot study using the same. Finally we present the refined tasks and briefly outline our large-scale experimental setup.

% \begin{enumerate}
% 	\item task design, temporal ir search needs, temporalia
% 	\item tasks for IIR - diane kelly
% 	\item studying search behavior - hugo, bota
% \end{enumerate}

% section background (end)
%\input{chapters/problem-definition}
\section{Search Task Creation} % (fold)
\label{sec:search_tasks}

\vspace{-4pt}
\begin{table*}[htpb]
\centering
\small
\begin{tabularx}{0.9\textwidth}{l|l|p{2cm}|p{6cm}}
\toprule
\textbf{Process} & \textbf{Target Outcome} & \textbf{Time} & \textbf{Post-Search Quiz}   \\ \midrule
Remember & Fact & 2 min. &  Fill-in-the-blank  \\ \hline
\multicolumn{3}{p{6cm}|}{When was the Internet Explorer browser released by Microsoft?} & Internet Explorer was released in the year $\rule{1cm}{0.15mm}$ \\ \midrule
\newline
Understand & List (set) & 10 min. &  True or False  \\ \hline
\multicolumn{3}{p{6cm}|}{Newt Gingrich was Speaker of the House in the United States Congress in the late 90s. Although he constantly battled President Clinton at the time, he won several accolades and spurred significant reforms. However he resigned from his post after just 4 years because his party lost faith in him. Why did this happen? } & {Newt Gingrich officially departed the House of Representatives in 1997 \newline
Newt Gingrich is a Republican. \newline
Newt Gingrich called for the United States federal government shutdown of 1995 and 1996. \newline
Newt Gingrich was ousted after an ethics violation \newline
Newt Gingrich was charged with running a politically biased TV advertisement.} \\ \midrule
\newline
Analyze & List (prioritized), Description & 15 min. &  Chronological Ordering \\ \hline
\multicolumn{3}{p{6cm}|}{Allen Iverson had a glittering career in the world of basketball. You are tasked with writing his biography and must chronicle his rise to stardom. Which teams did he play for both in college and the NBA? What were his major accolades in the NBA?} & {Iverson joins Georgetown \newline
Iverson wins MVP in the all star game and reaches the NBA playoff finals \newline
Iverson named NBA Rookie of the Year \newline
Iverson joins the Philadelphia 76ers \newline
Iverson joins the Denver Nuggets} \\ \bottomrule

\end{tabularx}
\caption{Examples of Search Tasks and Quizzes corresponding to Cognitive Processes and their Target Outcomes. All tasks and quizzes are available at~\texttt{\url{http://bit.ly/ArchiveSearchTasks}}.}
\label{tab:tasks}
\end{table*}

\subsection{Document Corpus} % (fold)
\label{ssub:document_corpus}

Unlike more general search behavior studies that focus on web search, we had to select an appropriate document corpus for our study. We selected the New York Times Annotated Corpus~\cite{nyt} which consists of 1.8 million news articles published between 1987 and 2007. The documents are stored in XML format and can be easily indexed. The documents are timestamped, explicitly categorized and also contain entity annotations.
%that are useful for retrieval models like~\cite{}.
% subsubsection document_corpus (end)

\subsection{Tasks} % (fold)
\label{sub:tasks}

When creating tasks, we kept two factors in mind: (i) the tasks must be relevant for the NYT corpus (ii) the tasks must be amenable to crowd sourcing. By amenable here we mean search tasks which negate the effect of a crowd worker's prior knowledge and the outcome of which can be easily evaluated at scale. We selected 3 domains: politics, sports and science and technology. Within each domain we carefully selected topics that are relatively unpopular for a global audience. For example, in sports (basketball) we chose Allen Iverson over Michael Jordan. We also selected only 3 levels of Krathwohl and Anderson's model: remember, understand and analyze. It was more challenging to design plausible tasks for the other levels when using news archives. We created 3 topics for each domain and complexity level resulting in a total of 27 search tasks. 

%Our intention of crowd sourcing such a study is to get high redundancy and a diverse user group for each experimental condition. We intend to use CrowdFlower whose demographics show a skew towards male asian adults. News search does not require domain knowledge when compared to searching legal case files. However it is necessary to create search tasks about topics that are relatively unpopular on a global level in order to negate the effect of prior knowledge.

%In our scenario situating the search tasks also becomes difficult. 
To motivate crowd workers and measure task performance, we intend to compensate workers for participation and based on their performance in a post-search quiz. This is different from lab based studies like~\cite{kellyictir} where participants are given fixed rewards and required to either copy-paste text from documents as evidence or write an explanation. Furthermore this requires manual judgment to grade performance whereas quizzes can be evaluated automatically. Table~\ref{tab:tasks} shows an example of a search task and corresponding quiz for each complexity level. The type of quiz was selected based on target outcomes suggested in~\cite{kellyictir}.
\begin{itemize}
	\item \textbf{Remember}: A remember search task entails the search for a single fact. We created search tasks where the user has to search for the date of a particular event. The post-search quiz is a fill-in-the-blank statement.
	\item \textbf{Understand}: Understanding tasks are more complex since they require the user to compile a list of facts from more than one document. We created tasks that needed the user to explain a certain incident in the past. All search tasks in this category ask the user ``Why did ...'', followed by the incident. The post-search quiz is a set of statements that the user has to decide are true or false. In our pilot study, we include a ``don't know'' option as well.
	\item \textbf{Analyze}: Analyze tasks force the user to not only assimilate information but also order the facts learned. We created search tasks where the goal was to write a biography for a given entity, ensuing that the user has to identify, compile and summarize facts. The post-search quiz was an order the facts challenge where users have to chronologically order a set of statements.
\end{itemize}

We further set an arbitrary time limit for each task in order to make the study tractable. All tasks are made available to the public for further research. Notice that the search tasks are entity centric. We selected entities that had at least 1000 articles published about them in our corpus. We created the quizzes by consulting Wikipedia and searching the corpus. While we are aware that some of our search tasks can be solved by simply looking up Wikipedia, our intention is to simulate the kind of tasks people would carry out when working with much older or specialized archives which are difficult to obtain.

% subsection tasks (end)

% \begin{enumerate}
% 	\item dataset
% 	\item choice of complexities, domains, table of topics
% 	\item how did you select the pre task and post task questions
% 	\item time limits
% \end{enumerate}

% section setup (end)

\section{Pilot Study} % (fold)
\label{sec:pilot_study}

We conducted a pilot study with 6 participants and a sample of 6 topics (2 from each category) to confirm if (a) the tasks are varied in complexity (b) the tasks are well defined and (c) the features and interface of the search engine are useful for these tasks.

\subsection{Temporal Search Engine} % (fold)
\label{sub:temporal_search_engine}
Each participant had access to the same search engine and interface. The SERP is shown in Figure~\ref{fig:ui}. Next to the query text field is a countdown timer informing participants of the time left. There is also a ``Done'' button to terminate the task early and start the post-search quiz and questionnaire. The timeline provides the user with visual cues for important time periods that she can refine her search with. The timeline is created by aggregating the top 1000 articles returned based on month of publication. We highlight important time periods using a simple burst detection algorithm. Users can interact with the timeline in 2 ways - click and drag across the timeline to filter by an arbitrary time interval or filter by directly clicking on a highlighted burst.
%All activities within the browser tab are recorded, including when the tab loses focus. 
Note that in this paper we do not wish to study the effects of the user interface but instead wish to check if the features are even used in any of the tasks. Hence we do not use variants of the search interface in the pilot study.

\subsection{Study Setup} % (fold)
\label{sub:study_setup}

We recruited 6 graduate students from our department to complete 3 search tasks each. All participants were given the same instructions explaining the user interface and the retrieval models. Participants completed their tasks remotely and were asked to send us an email with any feedback they may have. We used the pre and post search questionnaires from~\cite{kellyictir} to measure task difficulty, engagement and interest. We also appended the following questions to the post-search questionnaire: Was the timeline useful for this task? Was the ability to switch between rankings useful for this task? Was the time provided sufficient?

To check for prior knowledge in the pre-search test, participants were asked to rate how well they know the history of the entity in question (before they are shown the search task) on a 5 point Likert scale. For future studies, participants claiming to know the entity well will be assigned a different task. Participants were asked to perform the study on desktops or laptops only. They were shown a random task from each category in random order.

\subsection{Results and Discussion} % (fold)
\label{sub:results}

All users completed a task from each category. All users also reported low prior knowledge of the entity in question and also moderate interest in the search task which are desirable traits. We also asked participants how often they searched for news on the web at the start of the study. Only 1 participant claimed to have never searched for news while the majority searched once a week. The pre-search questionnaires showed that participants found remember tasks to be the most defined, followed by understand tasks and then analyze. We found the oppposite trend for the question ``How difficult do you think it will be to search for information for this task using this search engine?'', with remember tasks being considered the easiest. Interestingly all 3 types were reported to have a well defined outcome with no discernible difference in the responses. 

Participants were shown the post-search questionnaire only after completing the quiz. This allowed participants to objectively measure if they had performed well and then report experienced task difficulty and complexity. We found that all participants answered the quiz for remember tasks correctly but had average performance in other task types. The first result we noticed was that even though all tasks seemed well defined, users felt that it is was hard to determine when they had gathered enough information for analyze and understand tasks. Additionally in both task types, only one participant felt the time allotted was sufficient. Participants also found it difficult to decide if a document is relevant or not. This can be directly attributed to the nature of the corpus -- most news articles tend to be long and highly descriptive. They all usually report significant incidents and deciding which ones to read based on a snippet and title is harder than web search. Finally, from our logs we found that analyze tasks resulted in the highest number of interactions (clicks, queries, timeline usage), followed by understand and remember tasks (Table~\ref{tab:stats}). An interesting insight here was that users tended to use both retrieval models for the same query string in quick succession. 

\begin{table}[]
\centering
\small
\begin{tabular}{cccc}
\hline
  \textbf{Task} & \textbf{Queries}  &  \textbf{Timeline} &   \textbf{Clicks} \\ \hline
 Remember  & 3  & 1  &  2  \\ \hline
 Understand & 7  & 2  & 6   \\ \hline
 Analyze  &  13 & 7 &  11  \\ \hline
\end{tabular}
\caption{\small{Search Log Statistics from the Pilot Study. Values are averaged and rounded up across all participants. Timeline refers to the average number of times users chose to filter results using the timeline. Note that we consider a search string with the selected retrieval model as a new ``query''}}
\label{tab:stats}
\vspace{-10pt}
\end{table}

All participants also reported being highly engaged by the task. They also agreed that the timeline was between moderate to highly useful for all task types. The ability to switch retrieval models was considered useful for understanding and analyzing tasks and unnecessary for remember tasks. The proportion of queries using HistDiv was slightly less than $50\%$ in all cases. In their email correspondences, 3 participants felt that note taking would be a useful feature to add. Based on these results, we refined our search tasks and setup in the following way: (a) increased time limit for understanding and analyze tasks (b) indications regarding the type of post-search quiz to make it easier to decide when to stop (c) emphasizing the lead paragraph of a news article (which usually summarizes the article) when users decide to read a result.

% subsection results (end)

% subsection study_setup (end)

% subsection temporal_search_engine (end)

% \begin{enumerate}
% 	\item setup- number of participants, questionnaires, system variant
% 	\item overall time needed for the study
% 	\item post task, discussion
% \end{enumerate}

% section pilot_study (end)
%\input{chapters/results}

\section{Conclusion} % (fold)
\label{sec:conclusion}
To study search behavior and interfaces designed for archive search, we designed a set of search tasks based on directives from Kelly et al.~\cite{kellyictir}. The tasks are tailored to the New York Times archive which spans 20 years from 1987 to 2007. We also provide pre and post-search tests specific to each task and complexity level to make evaluation of user performance more scalable. Based on a pilot study, we found that the tasks have the desired effect of increasing difficulty and search activity. We identified weaknesses in our setup like the lack of time and hard-to-discern document relevance. The results show that the tasks and system, with minor modifications, are viable for a crowdsourcing based experiment to study user search behavior in archives.

\vspace{25pt}
\section{Acknowledgments}

This work is funded by the ERC
under the grant 339233 (ALEXANDRIA).

% section conclusion (end)

%\input{chapters/measures}
%\input{chapters/retrievalmodel}

% \section{Acknowledgments}
% This section is optional; it is a location for you
% to acknowledge grants, funding, editing assistance and
% what have you.  In the present case, for example, the
% authors would like to thank Gerald Murray of ACM for
% his help in codifying this \textit{Author's Guide}
% and the \textbf{.cls} and \textbf{.tex} files that it describes.
{ \footnotesize
\bibliographystyle{abbrv} 
\bibliography{bib/chiir.bib}  % sigproc.bib is the name of the Bibliography in this case

\begin{thebibliography}{10}

\bibitem{ukarchive}
British newspaper archive http://www.britishnewspaperarchive.co.uk/.

\bibitem{neat}
O.~Alonso, K.~Berberich, S.~Bedathur, and G.~Weikum.
\newblock Neat: News exploration along time.
\newblock In {\em Proceedings of ECIR}, 2010.

\bibitem{tir}
O.~Alonso, J.~Str{\"o}tgen, R.~A. Baeza-Yates, and M.~Gertz.
\newblock Temporal information retrieval: Challenges and opportunities.
\newblock {\em TWAW}, 2011.

\bibitem{Anand:2011}
A.~Anand, S.~Bedathur, K.~Berberich, and R.~Schenkel.
\newblock Temporal index sharding for space-time efficiency in archive search.
\newblock In {\em Proceedings of ACM SIGIR}, 2011.

\bibitem{berberich2010language}
K.~Berberich, S.~Bedathur, O.~Alonso, and G.~Weikum.
\newblock A language modeling approach for temporal information needs.
\newblock In {\em Proceedings of ECIR}, 2010.

\bibitem{braun2002information}
L.~Braun, F.~Wiesman, and I.~Sprinkhuizen-Kuyper.
\newblock Information retrieval from historical corpora.
\newblock In {\em Proceedings of the DIR}, pages 106--112. Citeseer, 2002.

\bibitem{campos_survey_2014}
R.~Campos, G.~Dias, A.~M. Jorge, and A.~Jatowt.
\newblock Survey of temporal information retrieval and related applications.
\newblock {\em {ACM} Comput. Surv.}, 47(2), 2014.

\bibitem{Gupta2014}
D.~Gupta and K.~Berberich.
\newblock Identifying time intervals of interest to queries.
\newblock In {\em Proceedings of the 23rd ACM CIKM}, 2014.

\bibitem{jarvelin2015information}
A.~J{\"a}rvelin, H.~Keskustalo, E.~Sormunen, M.~Saastamoinen, and K.~Kettunen.
\newblock Information retrieval from historical newspaper collections in highly
  inflectional languages: A query expansion approach.
\newblock {\em Journal of the Association for Information Science and
  Technology}, 2015.

\bibitem{kellyictir}
D.~Kelly, J.~Arguello, A.~Edwards, and W.~Wu.
\newblock Development and evaluation of search tasks for iir experiments using
  a cognitive complexity framework.
\newblock In {\em Proceedings of ACM ICTIR}, 2015.

\bibitem{krathwohl}
D.~R. Krathwohl.
\newblock A revision of bloom's taxonomy: An overview.
\newblock {\em Theory into practice}, 41(4), 2002.

\bibitem{croft}
X.~Li and W.~B. Croft.
\newblock Time-based language models.
\newblock In {\em Proceedings of ACM CIKM}, 2003.

\bibitem{liu2010search}
J.~Liu, M.~J. Cole, C.~Liu, R.~Bierig, J.~Gwizdka, N.~J. Belkin, J.~Zhang, and
  X.~Zhang.
\newblock Search behaviors in different task types.
\newblock In {\em Proceedings of the Conference on Digital libraries}. ACM,
  2010.

\bibitem{Maslov:2006:ENQ:1135777.1135950}
M.~Maslov, A.~Golovko, I.~Segalovich, and P.~Braslavski.
\newblock Extracting news-related queries from web query log.
\newblock In {\em Proceedings of the 15th International Conference on World
  Wide Web}, WWW '06, 2006.

\bibitem{mishra2015expose}
A.~Mishra and K.~Berberich.
\newblock Expos{\'e}: Exploring past news for seminal events.
\newblock In {\em Proceedings of WWW}. ACM, 2015.

\bibitem{wildemuth}
P.~Peter~Willett, B.~Wildemuth, L.~Freund, and E.~G.~Toms.
\newblock Untangling search task complexity and difficulty in the context of
  interactive information retrieval studies.
\newblock {\em Journal of Documentation}, 70(6):1118--1140, 2014.

\bibitem{nyt}
E.~Sandhaus.
\newblock The new york times annotated corpus.
\newblock {\em Linguistic Data Consortium, Philadelphia}, 6(12):e26752, 2008.

\bibitem{expedition}
J.~Singh, W.~Nejdl, and A.~Anand.
\newblock Expedition: A time-aware exploratory search system designed for
  scholars.
\newblock In {\em Proceedings of the ACM SIGIR}, 2016.

\bibitem{histdiv}
J.~Singh, W.~Nejdl, and A.~Anand.
\newblock History by diversity: Helping historians search news archives.
\newblock In {\em Proceedings of the ACM CHIIR}, 2016.

\end{thebibliography}

}

\end{document}